\documentclass{webofc}
\usepackage[varg]{txfonts}   
\begin{document}
\title{Heavy-flavor meson and baryon production in high-energy nucleus-nucleus collisions}
%
%

\author{\firstname{A.} \lastname{Beraudo}\inst{1}\fnsep\thanks{\email{beraudo@to.infn.it}} \and
        \firstname{A.} \lastname{De Pace}\inst{1}\fnsep \and
        \firstname{D.} \lastname{Pablos Alfonso}\inst{1}\fnsep \and
        \firstname{F.} \lastname{Prino}\inst{1}\fnsep \and 
        \firstname{M.} \lastname{Monteno}\inst{1}\fnsep \and
        \firstname{M.} \lastname{Nardi}\inst{1}}

\institute{INFN - Sezione di Torino, via Pietro Giuria 1, I-10125 Torino}

\abstract{We present a new model for the description of heavy-flavor hadronization in relativistic heavy-ion collisions. We explore its effect on the charmed hadron yields and kinematic distributions once the latter is applied at the end of transport calculations used to simulate the propagation of heavy quarks in the deconfined fireball produced in the collision. The model is based on the formation of color-singlet clusters through the recombination of charm quarks with light antiquarks or diquarks from the same fluid cell. This local mechanism of color neutralization leads to a strong space-momentum correlation, which provides a substantial enhancement of charmed baryon production and of the collective flow of all charmed hadrons.}
\maketitle
\section{Introduction}\label{Sec:intro}
The diffusion of Brownian particles can be used to access microscopic properties of the medium in which their propagation takes place. Historically, the diffusion of small grains in water was used by Perrin to prove the discontinuous structure of matter and to provide a first estimate of the Avogadro number, finding ${\cal N}_A\approx 5.5-7.2\cdot 10^{23}$~\cite{Perrin:1926}. More than one century later one of the goals of heavy-ion collisions is to exploit the \emph{relativistic} brownian motion of heavy quarks to get an estimate of comparable accuracy of medium properties like the \emph{momentum broadening coefficient} $\kappa$. An important difference in this case is that the nature of the Brownian particle changes during its propagation, since after a few fm/c it undergoes hadronization. This introduces a source of systematic uncertainty in the extraction of transport coefficients; on the other hand it can be considered an issue of interest in itself, in particular to study how hadronization changes in the presence of a medium acting as a color reservoir. This, in the case of charm quarks, will be the subject of our study.
\section{The model}\label{Sec:model}
\begin{figure*}
\centering
\includegraphics[clip,width=0.48\textwidth]{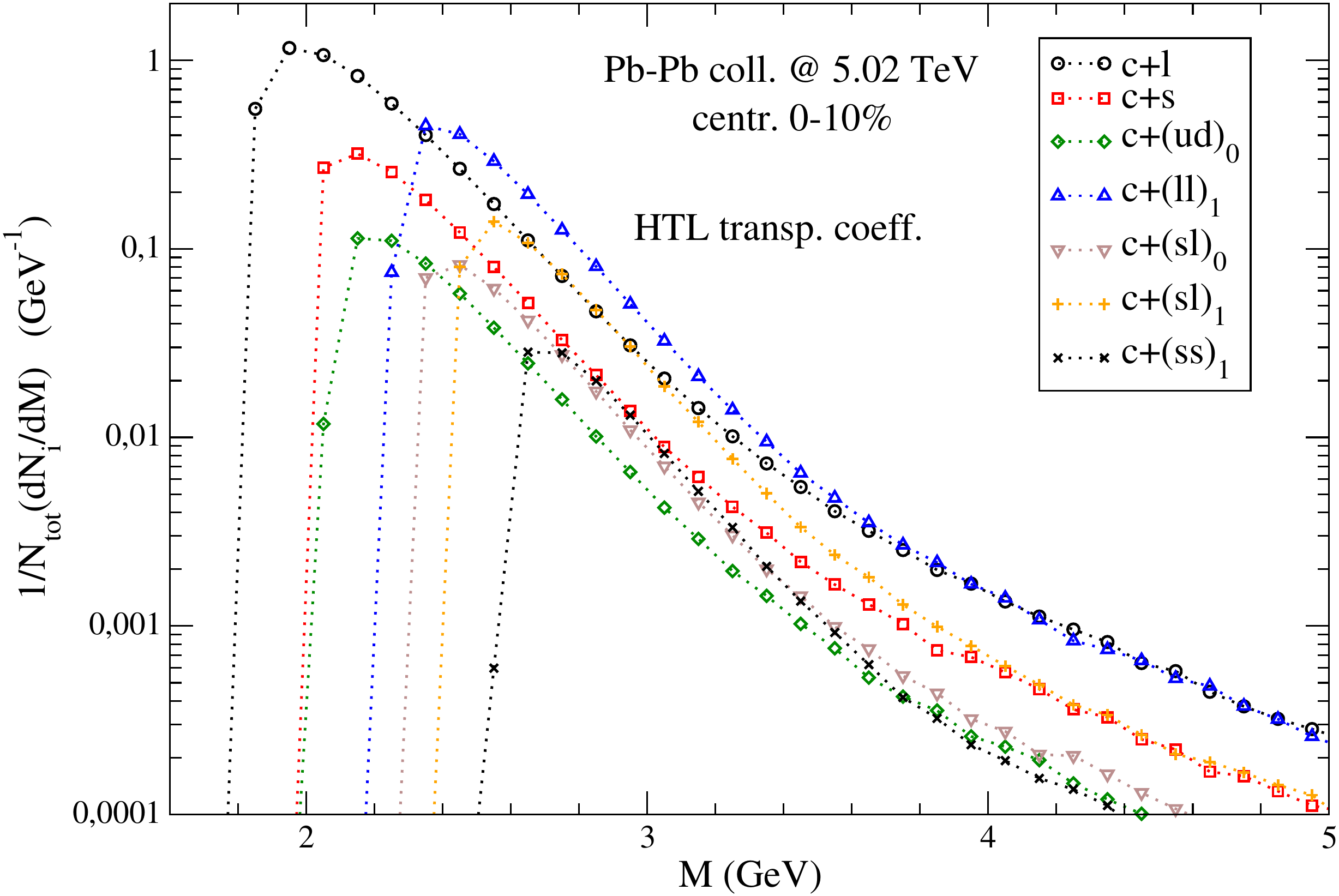}
\includegraphics[clip,width=0.48\textwidth]{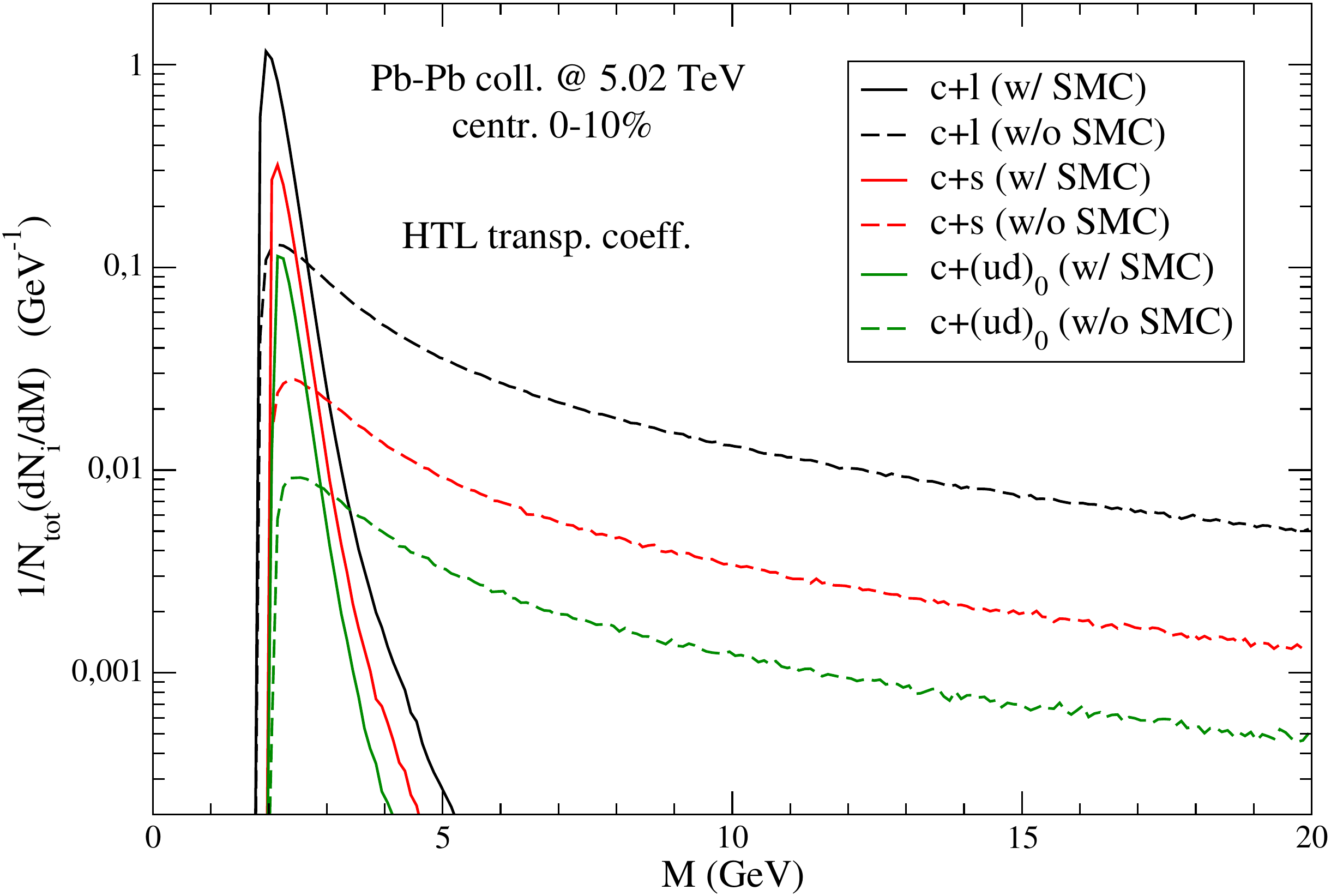}
\caption{Left panel: the invariant-mass distribution of the $Q\overline q$ and $Q(qq)$ clusters at the hadronization hypersurface in the different isospin, strangeness and spin channels. Right panel: cluster invariant-mass distributions in case recombination occurs locally (continuous curves, default implementation), with strong correlation between the parton momenta and position, or non-locally (dashed curves), with no space-momentum correlation. All results refer to central Pb-Pb collisions at $\sqrt{s_{\rm NN}}\!=\!5.02$ TeV, with charm quark propagation before hadronization described by HTL transport coefficients.}
\label{fig:Mdistr}       
\end{figure*}
Any hadronization model must start grouping colored partons into color-singlet structures which will give rise to the final hadrons. In ``elementary'' hadronic collisions partons are taken from the hard process, from the shower stage, from the underlying event and from the beam remnants. In heavy-ion collisions recombining partons are taken from the hot medium produced in the collision and they are \emph{close in space}: since the fireball undergoes a collective expansion, this last detail has deep phenomenological consequences. 

We now briefly describe our model for in-medium charm hadronization. For more details we refer the reader to our original publication~\cite{Beraudo:2022dpz}.
Once a $c$ quark, during its stochastic propagation through the fireball, reaches a fluid cell at $T_H=155$ MeV it recombines with a
light antiquark or diquark from the same fluid element. Both the species and the momentum of the medium particle are sampled assuming they obey a thermal distribution in the local rest frame (LRF) of the fluid. The thermal particle is then boosted to the laboratory frame and recombined with the charm quark, leading to the formation of the cluster ${\cal C}$. Since the $c$ quark and the thermal particle are taken from the same fluid cell typically their momenta are quite collinear (Space-Momentum Correlation) and this leads to the production of quite low invariant-mass clusters,  
as one can see from the left panel of Fig.~(\ref{fig:Mdistr}). We also check that, suppressing SMC's by randomly redistributing the HQ's on the decoupling hypersurface (see right panel of Fig.~\ref{fig:Mdistr}), the production of heavier clusters is favored.
As in Herwig~\cite{Webber:1983if}, one treats differently light and heavy clusters. Clusters with $M_{\cal C}<M_{\rm max}\approx 4$ GeV undergo an isotropic two-body decay in their LRF, giving rise to a charmed hadron and a second soft particle; the very rare heavier clusters are fragmented into multiple hadrons as Lund strings~\cite{Andersson:1983ia}.
\section{Results and discussion}\label{sec:results}
\begin{figure*}
\centering
        \includegraphics[clip,width=0.98\textwidth]{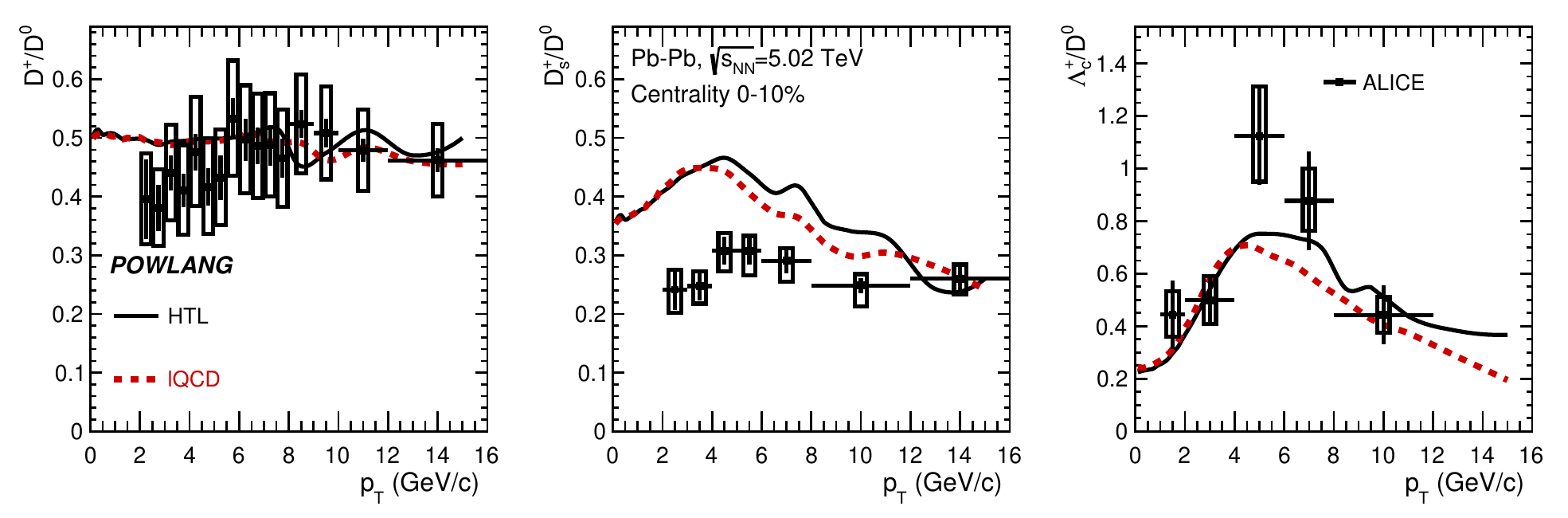}
\caption{Predictions for the relative yields of charmed hadrons (relative to $D^0$ mesons) in central Pb-Pb collisions at $\sqrt{s_{\rm NN}}\!=\!5.02$ TeV for different transport coefficients compared to recent ALICE data~\cite{ALICE:2021bib,ALICE:2021kfc,ALICE:2021rxa}.}\label{Fig:D-Ds-Lc}
\end{figure*}

\begin{figure*}
\centering
        \includegraphics[clip,height=5cm]{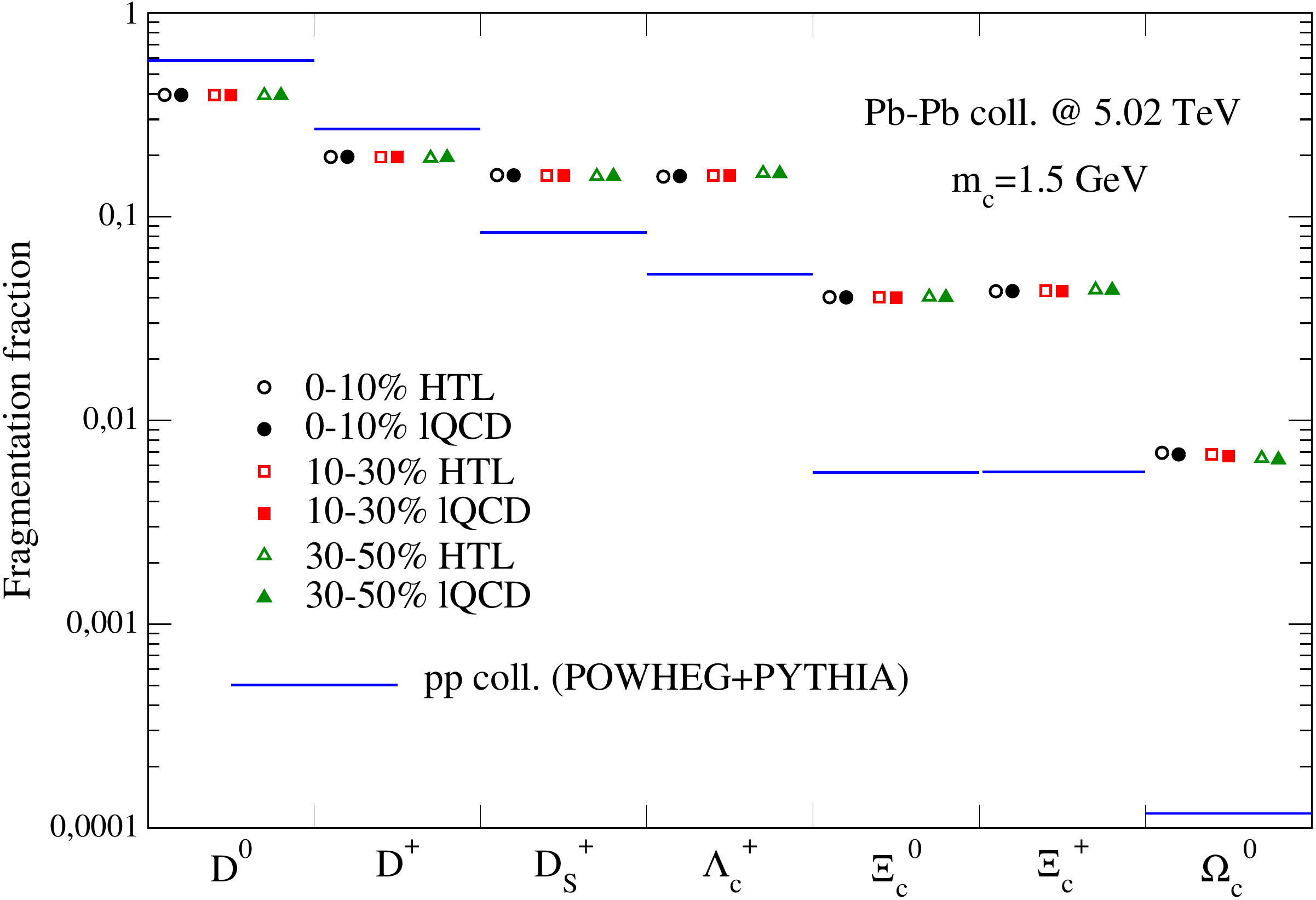}
        \includegraphics[clip,height=5cm]{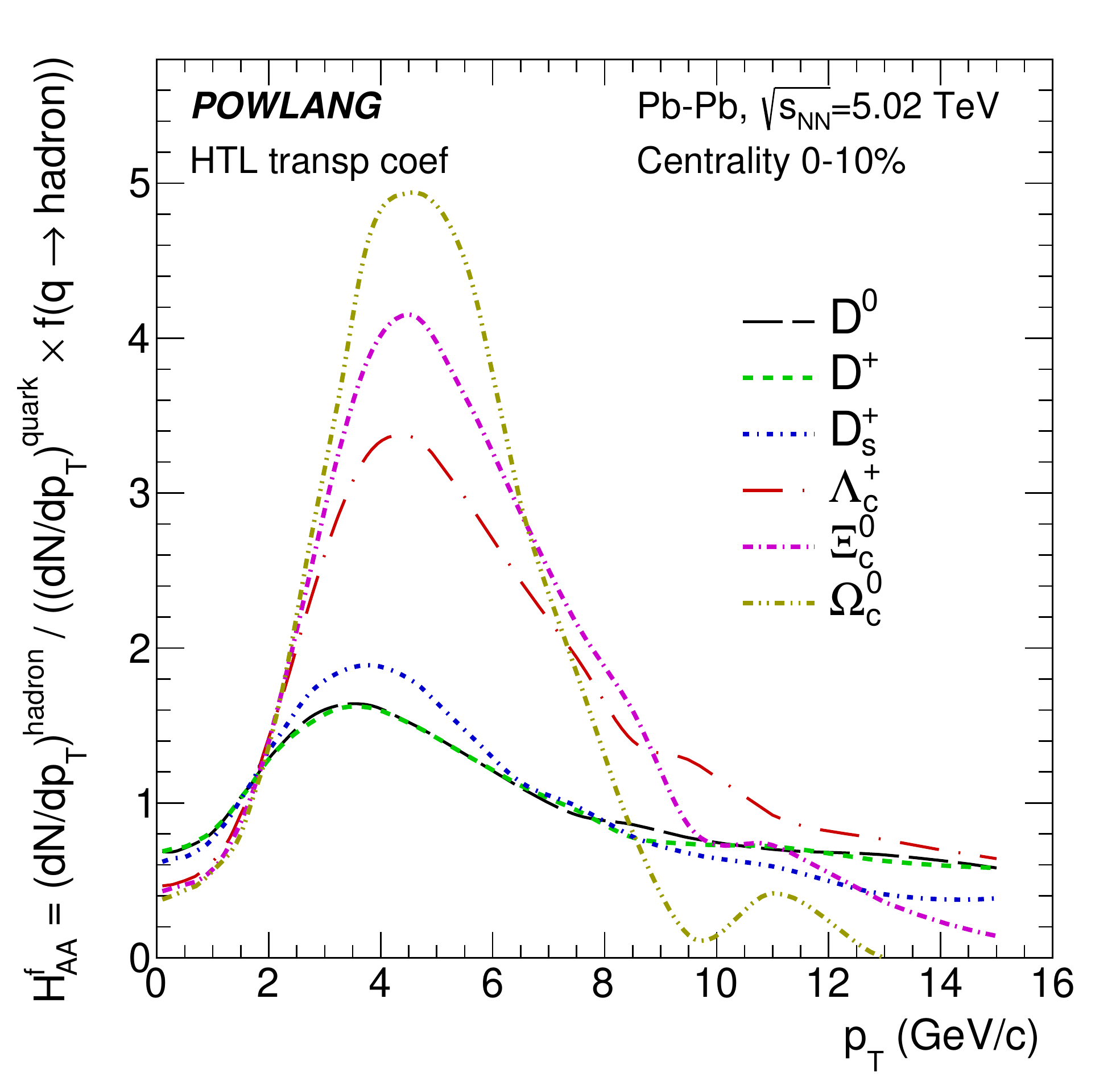}
\caption{Left panel: charm fragmentation fractions in Pb-Pb collisions for different centrality classes and transport coefficients; also shown, in blue, the results obtained for proton-proton collisions at the same center-of-mass energy obtained with the POWHEG-BOX package~\cite{Alioli:2010xd}. Right panel: hadron-to-quark ratio of the normalized $p_T$-distributions for the different species of charmed-hadrons.}\label{fig:hc-species}
\end{figure*}

Interfacing our new hadronization model to numerical simulations of heavy-quark transport in the deconfined fireball one can obtain a satisfactory description of important and partially unexpected experimental findings, in particular the strong enhancement of charmed-baryon production recently observed in heavy-ion~\cite{ALICE:2021bib} and also in proton-proton collisions~\cite{ALICE:2020wfu}, which cannot be explained by any hadronization model tuned to reproduce $e^+e^-$ data.  
An example of our results is given in Fig.~\ref{Fig:D-Ds-Lc}, where we plot as a function of $p_T$ the $D^+/D^0$, $D_s^+/D^0$ and $\Lambda_c^+/D^0$ ratios in Pb-Pb collisions at $\sqrt{s_{\rm NN}}=5.02$ TeV, comparing our predictions to recent ALICE data~\cite{ALICE:2021bib,ALICE:2021kfc,ALICE:2021rxa}. Notice that if hadronization occurred as in $e^+e^-$ collisions one would get $\Lambda_c^+/D^0\approx 0.1$. As one can see the strong enhancement of the $\Lambda_c^+/D^0$ ratio for intermediate values of $p_T$ is correctly reproduced. Analogous results are obtained when our model is compared to STAR data for Au-Au collisions at $\sqrt{s_{\rm NN}}=200$ GeV~\cite{STAR:2019ank}.
Integrating the momentum distributions one gets the \emph{fragmentation fractions} into the different charmed hadrons, which we display in the left panel of Fig.~\ref{fig:hc-species}, comparing them to the predictions for proton-proton collisions provided by a pQCD event generator~\cite{Alioli:2010xd}. Our results in Pb-Pb collisions are pretty independent from the centrality class and from the heavy-quark transport coefficients. The most relevant prediction is the huge enhancement of the production of all charmed baryons, in agreement with recent experimental observations.
In the right panel of Fig.~\ref{fig:hc-species} we show how hadronization, beside the yields, affects the kinematic distributions. To this purpose we plot the ratio of the $p_T$ spectra of the different charmed hadrons (normalized to the corresponding fragmentation fractions) relative to the one of the parent quarks. As one can see, for all hadron species such a ratio displays a bump at intermediate values of $p_T$, the enhancement being particularly relevant for charmed baryons and increasing with the mass of the particle.

\begin{figure*}
\centering
        \includegraphics[clip,height=5cm]{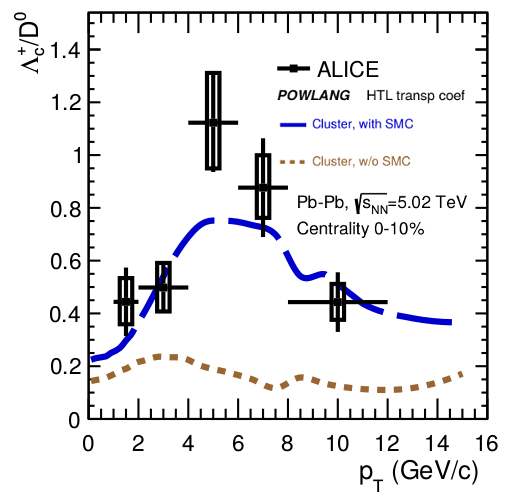}
        \includegraphics[clip,height=5cm]{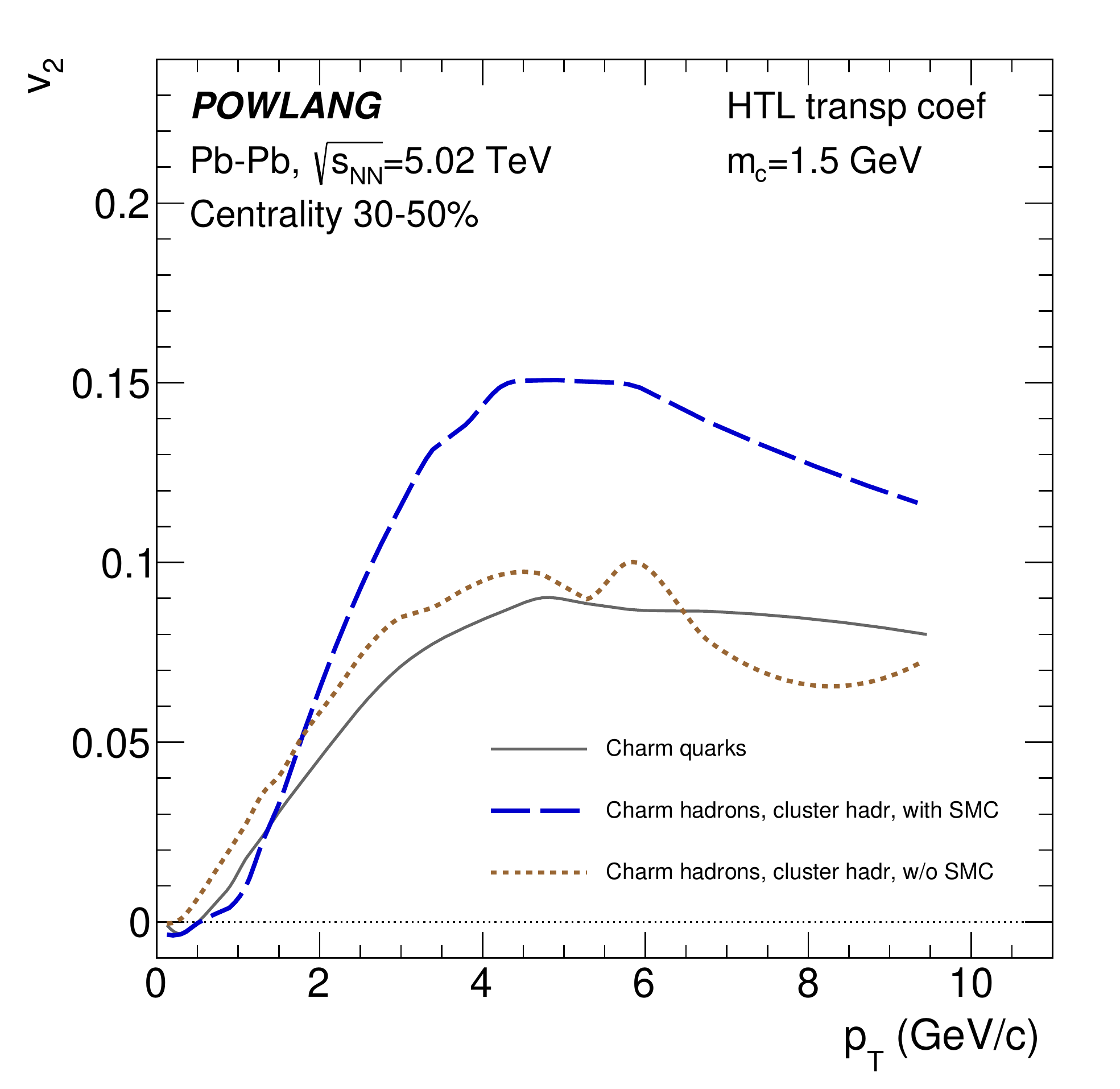}
\caption{Charmed hadron ratios (left panel) and elliptic flow (right panel) with (dashed blue curves) and without (dotted brown curves) space-momentum correlation, the last case being obtained mixing momentum and position of heavy quarks from different events at hadronization.}\label{Fig:yields-noSMC}
\end{figure*}

The above results can be interpreted as follows within our model. First of all the possibility for $c$ quarks to recombine with a light diquark enhances the fraction of clusters carrying baryon number $B\!=\!1$, which will decay into a charmed baryon. Secondly, SMC favors the recombination of $c$ quarks with a thermal particle with a quite collinear momentum, parallel to the velocity of the fluid cell. Thus, the resulting clusters typically have a quite low invariant mass and, furthermore, receive a boost due to the collective flow of the fireball. The effect is particularly important for the heaviest diquarks, due to the mass dependence of the radial flow. Furthermore, the light invariant mass of most clusters usually leaves room only for a two-body decay, in which the daughter charmed hadron carries a sizable momentum fraction of the ancestor. As a result, hadronization via local recombination is an efficient mechanism to transfer the collective flow of the fireball to the final charmed hadrons and this holds not only for their $p_T$-distributions, but also for their azimuthal anisotropies, as can be appreciated from the right panel of Fig.~\ref{Fig:yields-noSMC}).      

The importance of SMC's can be accessed by artificially breaking the connection between the position and the momentum of the hadronizing particles. This can be implemented by randomly redistributing the quarks undergoing recombination over the hadronization hypersurface. The effect on the final charmed hadrons is shown in Fig.~\ref{Fig:yields-noSMC}: the $\Lambda_c^+/D^0$ ratio is no longer enhanced, but one finds a value around 0.1, as in $e^+e^-$ collisions; furthermore the charmed hadron $v_2$ is lower, very similar to the one of the parent heavy quarks. Recombining partons are no longer collinear: this suppresses the collective flow of the clusters and increases their invariant mass, leading them to hadronize via string fragmentation, with no enhanced baryon production.

\bibliography{SQM-proce}

\end{document}